\documentclass[aps,prb,reprint]{revtex4-2}
\usepackage{amsmath}
\usepackage{amssymb}
\usepackage{graphicx}
\usepackage{float}
\usepackage[colorlinks=true,linkcolor=blue,urlcolor=blue,citecolor=blue]{hyperref}
\newcommand{\degree}{^\circ}

\begin{document}
\title{Ferroelasticity tunable altermagnets}
\author{Ning Ding}
\author{Haoshen Ye}
\author{Shan-Shan Wang}
\email{wangss@seu.edu.cn}
\author{Shuai Dong}
\email{sdong@seu.edu.cn}
\affiliation{Key Laboratory of Quantum Materials and Devices of Ministry of Education, School of Physics, Southeast University, Nanjing 21189, China}
\date{\today}
\begin{abstract}
Altermagnets have garnered great interest due to their non-relativistic spin splitting and novel physical properties. However, the control of altermagnetic states remains underexplored. Here, we propose a unique multiferroic state, i.e. ferroelastic altermagnetic state, in which ferroelastic strain couples directly to the spin-splitting. Through symmetry analysis and first-principles calculations, we identify the ferroelastic $d$-wave altermagnetism of puckered pentagonal CoSe$_2$ monolayer. Interestingly, uniaxial stress can induce a ferroelastic phase transition, accompanied by a $90\degree$ rotation of the spin-splitting bands. Cooperative rotation of the lattice and N\'eel vectors preserves the sign of Kerr angle, whereas noncooperative rotation reverses it. Our work provides a general strategy for manipulating altermagnetism in multiferroic systems and opens other avenues for exploring emergent magnetoelastic phenomena.  
\end{abstract}
\maketitle

Altermagnetism represents an emerging research frontier of condensed matter research, which hosts unconventional magnetic phenomena and harnesses their unique physical properties for technological innovation~\cite{Libor2022-PRX-9,Libor2022-PRX-12,Ma2021-NC,Bai2024-AFM}. With antiferromagnetic spin configurations, altermagnets exhibit non-relativistic spin splitting in Brillouin zone, and thus combine the advantageous features of both ferromagnets and antiferromagnets. Specifically, they can display a magneto-optical response similar to ferromagnets while maintaining the negligible stray fields and field robustness of antiferromagnets~\cite{Song2025-NRM,Gonzalez2023-PRL}. 

Recent experimental researches have discovered various altermagnetic materials, including MnTe~\cite{Krempasky2024-Nature,Lee2024-PRL,Hariki2024-PRL}, CrSb~\cite{Ding2024-PRL,Zeng2024-AS,Zhou2025-Nature}, $\alpha$-MnTe$_2$~\cite{Zhu2024-Nature}, KV$_2$Se$_2$O~\cite{Jiang2025-NP}, Mn$_5$Si$_3$~\cite{Reichlova2024-NC}, and CoNb$_4$Se$_8$~\cite{Regmi2025-NC}. Moreover, many theoretical works have predicted the existence of altermagnetism in many van der Waals systems~\cite{Sodequist2024-APL,Jin2024-PRB,Milivojevic2024-2D-Mater,Chen2025-JACS-Au,Li2025-NL,Zhu2024-NL,Wang2025-NL,Tan2025-PRB}. Stacking and twisting engineering has been proposed to be a promising route to produce two-dimensional (2D) altermagnets~\cite{He2023-PRL,Liu2024-PRL,Pan2024-PRL,Sun2024-PRB,Zeng2024-PRB}. 

Multiferroic systems combining more than one ferroic orders, usually magnetic and ferroelectric orders, hold promise for controlling magnetism through nonmagnetic fields~\cite{Dong2015-AP}. Inspired by this prospect, magnetoelectric coupling has been proposed in various ferroelectric/antiferroelectric altermagnets~\cite{Duan2025-PRL,Gu2025-PRL,Zhu2025-NL,Sun2024-NL,Sun2025-AM,guo2023-arxiv,Khan2025-MH,wangshuaiyu2025-arxiv,smejkal2024-arxiv,yang2025-arxiv,zhu2025-sliding-arxiv,Urru2025-PRB}, where ferroelectricity can be used to control altermagnetism, offering opportunities for electrically tunable spintronics. Electric field control is indeed experimentally straightforward to implement. However, this approach becomes ineffective when the system exhibits metallic behavior. Nevertheless, the metallic nature of altermagnetic systems enables promising functionalities such as spin-polarized current generation, prompting the need to develop alternative manipulation strategies for manipulating altermagnetism. In fact, multiferroics are not limited to ferroelectric magnets. Ferroelasticity is also a kind of ferroicity, broader than ferroelectricity~\cite{Wang2022-PRB,xu2021-JMCC,zhang2021-JMCC}, which can be coupled with altermagnets. This coupling could enable the control of spin-related properties such as the magneto-optical Kerr effect (MOKE) through ferroelastic strain engineering. However, research on the coupling between altermagnetism and ferroelasticity remains rare and urgently needed. 

\begin{figure}
	\centering
	\includegraphics[width=0.47\textwidth]{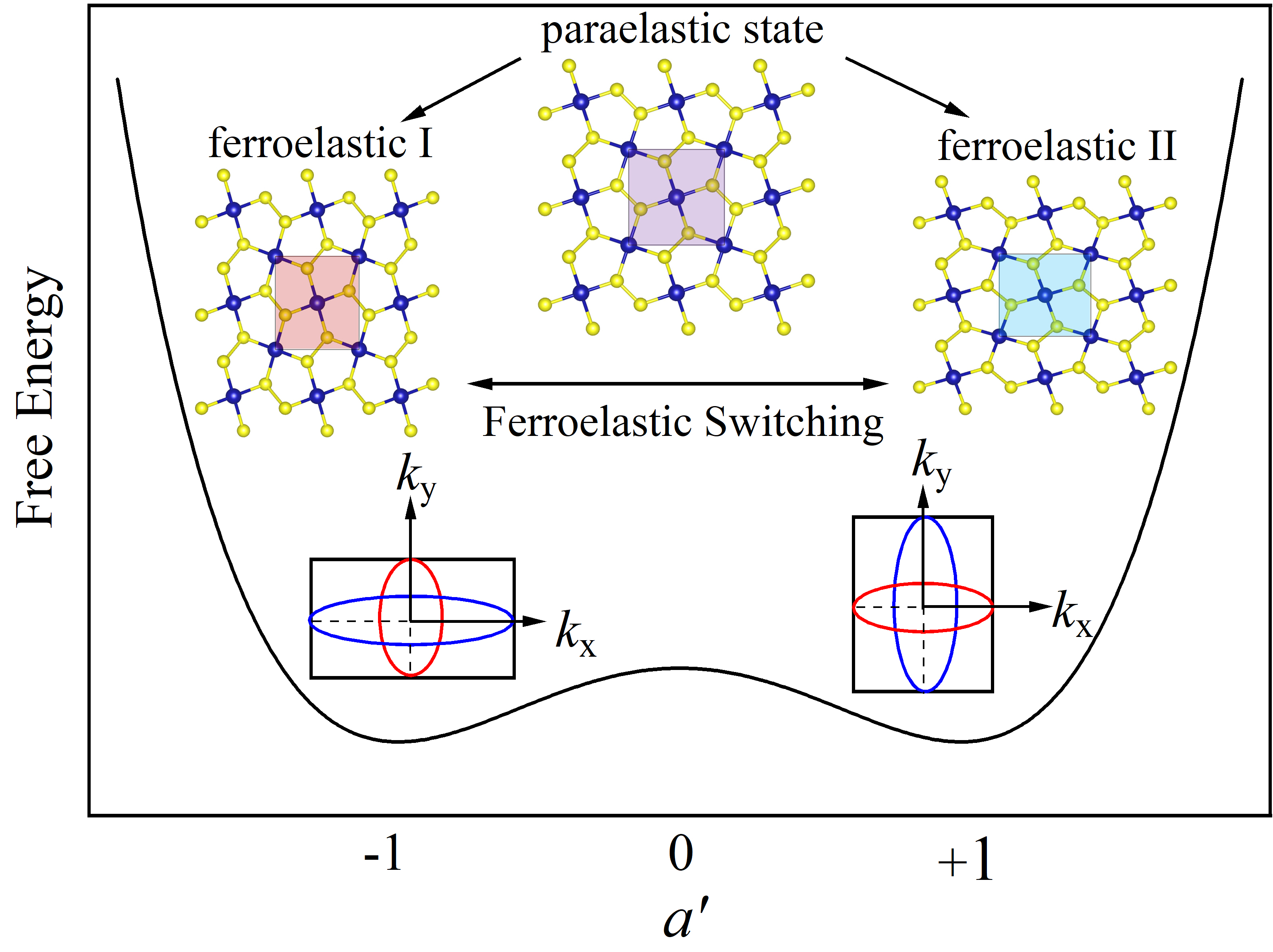}
	\caption{Schematic illustration of ferroelastic altermagnetism and its switching. We postulate the lattice constants ($a_0$ and $b_0$) of the ground state of the ferroelastic system, with $a_0<b_0$. The horizontal coordinate is $a'=(2a-a_0-b_0)/(b_0-a_0)$, where $a$ is lattice constant of $a$-direction during the ferroelastic switching. To enhance the visual clarity, the unit cells of ferroelastic states are represented by transparent colored boxes: light red denotes ferroelastic state I with lattice constant $a=a_0$ ($a'=-1$), purple represents the paraelastic state with $a=(a_0+b_0)/2$ ($a'=0$), and light blue indicates ferroelastic state II with $a=b_0$ ($a'=+1$). The following diagrams are spin-momentum locking diagrams of two ferroelasctic states.} 
	\label{fig1}
\end{figure}

In this Letter, we theoretically study the ferroelastic altermagnets, which enable ferroelastic control of spin splitting in electronic bands, as depicted in Fig.~\ref{fig1}. The material candidate is CoSe$_2$ monolayer, a 2D pentagonal lattice with unequal lattice parameters~\cite{Wang2025-AFM,Liu2018-PRB}. Connected by rotation symmetry, the opposite spin sublattices violate the parity-time reversal symmetry, and the time-reversal after fractional translation. As a result, the spin splitting occurs, which can be rotated by $90\degree$ through ferroelastic switching.

\textit{Methods}.
First-principles calculations were performed with the projector augmented-wave (PAW) potentials as implemented in Vienna $ab$ $initio$ Simulation Package (VASP)~\cite{PAW-PRB-1996}. The Perdew-Burke-Ernzerhof (PBE) parameterization of the generalized gradient approximation (GGA) was used for the exchange correlation functional~\cite{PBE-PRL-1996}. The Hubbard $U$ ($U_{\rm eff}=3$ eV) was applied using the Dudarev parametrization to better describe the correlation effects in transition-metal magnetic ions~\cite{Dudarev1998-PRB} according to the previous works~\cite{Liu2018-PRB,Wang2024-JPCC}. In addition, the cases of different Hubbard $U_{\rm eff}$ were shown in Fig.~S1 in the Supplemental Materials (SM)~\cite{sm}. A vacuum layer thicker than $15$ \AA\ was used to avoid artificial interactions from periodic boundary conditions. The plane-wave cutoff energy was $500$ eV, and Co's $4s3d3p$ electrons were treated as valence states. The $k$-point grid of $6 \times 6 \times 1$ was employed for structural relaxation and static calculations. The energy convergence criterion was set to $10^{-6}$ eV and the Hellman-Feynman forces criterion during structural relaxation was $0.005$ eV/\AA. The calculation of the phonon spectrum was used by the finite-difference method~\cite{Finite-differences-PRB-2002,Finite-difference-PRB-2005}. $Ab$ $initio$ molecular dynamics (AIMD) simulations were performed at $300$ K by using an $NVT$ ensemble that lasts $10$ ps with a time step of $4$ fs~\cite{AIMD}.

The tight-binding Hamiltonian, which incorporates Co-$d$ and Se-$p$ orbitals was constructed using maximally localized Wannier functions (MLWF) generated by the WANNIER90 code~\cite{Pizzi2020-JPCM}. The spin-resolved conductivity $\sigma_{ij}^{s}$ was calculated by using the BoltzWann module with a $k$-point mesh of $500 \times 500 \times 1$~\cite{BoltzWann}. The optical conductivity was calculated within the MLWF basis employing a high-density $k$-point grid of $200 \times 200 \times 1$~\cite{Di2010-RMP,Yao2004-PRL,Wang2006-PRB}.

CoSe$_2$ monolayer has the weak anisotropy, which can be confirmed by the nearly equal optical conductivity tensors ($\xi_\text{xx}$ and $\xi_\text{yy}$) as shown in Fig.~S2~\cite{sm}. Thereby, the MOKE can be described by the Kerr angle $\theta_{K}$ and ellipticity $\eta_{K}$, which can be quantitatively characterized by the following equation~\cite{Yang2021-PRB}:
\begin{equation}
\Phi_K=\theta_K+i\eta_K=i\frac{2\omega D}{c}\frac{\xi_\text{xy}}{\xi_\text{xx}^{sub}},
\end{equation} 
where $c$ represents the speed of light in vacuum, $\omega$ is frequency of incident light, and $D$ is the thickness of calculated material ($D=3$ \AA\ in our system). $\xi_\text{xx}^{sub}$ is the diagonal element of the optical conductivity tensor for SiO$_2$ substrate, which equals $i(1-n^2)\omega/4\pi$ with refractive index $n=1.546$.

To determine the magnetic ground state, the Monte Carlo (MC) method with the Metropolis algorithm was employed. A $20 \times 20$ lattice with periodic boundary conditions was used in the MC simulation, and larger lattices were also tested to verify the physical results. The initial $5 \times 10^5$ MC steps were discarded for thermal equilibrium, followed by $5 \times 10^5$ MC steps used for statistical averaging at each temperature. Temperature scanning was performed by using a quenching process.

\begin{figure}
	\centering
	\includegraphics[width=0.47\textwidth]{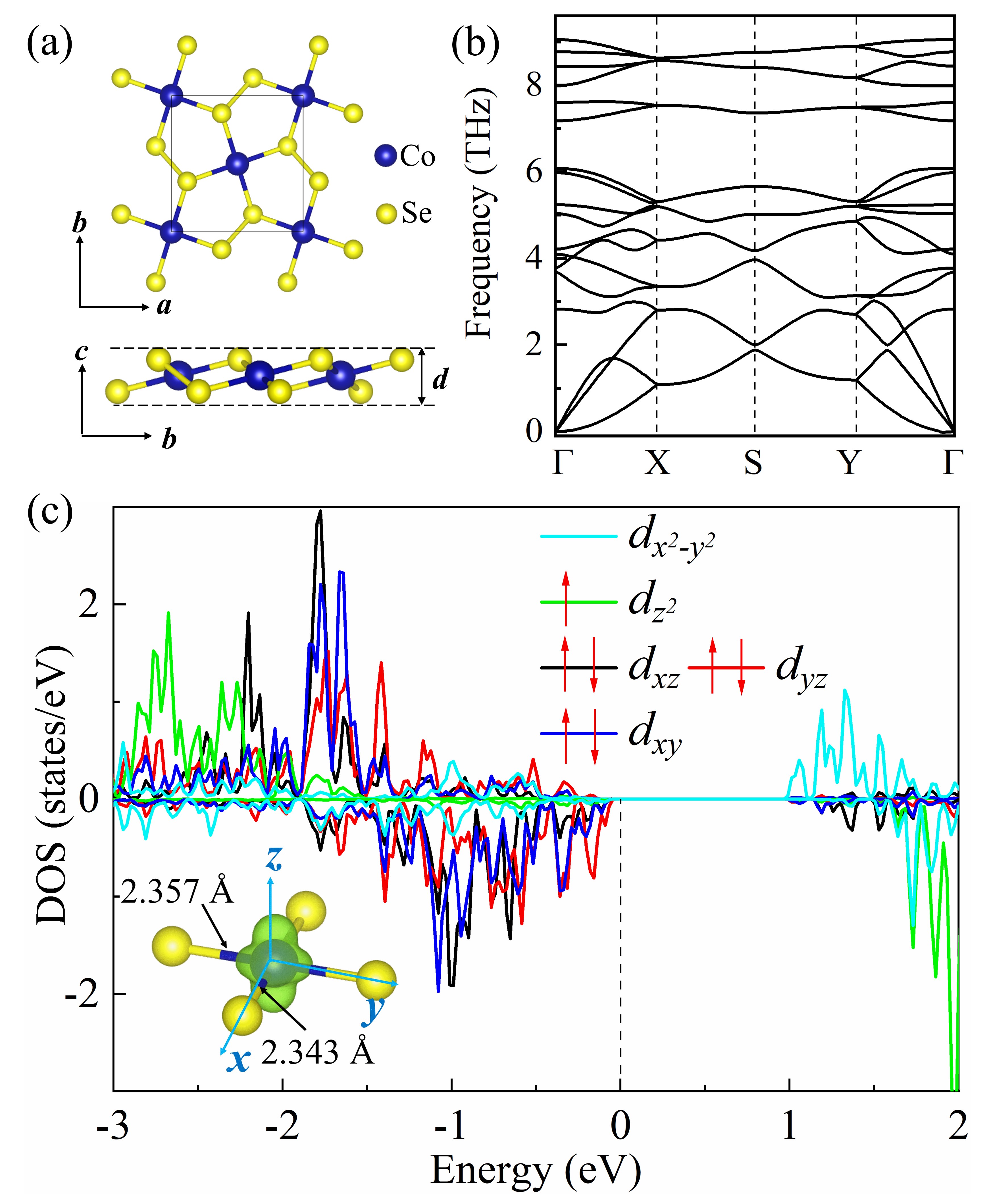}
	\caption{(a) Top and side views of pentagonal CoSe$_2$ monolayer. $d$: the thickness. (b) Phonon spectrum of monolayer CoSe$_2$. (c) Orbital-projected density of state (PDOS) of one spin-up Co ion. Inset: schematic of electron occupation of Co$^{2+}$'s $3d$ orbitals under a quasi-tetragonal crystal field. The unpaired $d_{z^2}$ orbital can be visualized by the spin polarized electron cloud colored in green.}
	\label{fig2}
\end{figure}

\textit{Structure and magnetism}.
The space group of pentagonal CoSe$_2$ monolayer is $P$2$_1$/$c$ (No. 14), identical to the previously predicted structural analogue CoS$_2$ monolayer~\cite{Liu2018-PRB}. As shown in Fig.~\ref{fig2}(a), CoSe$_2$ monolayer exhibits a buckled pentagonal structure. The buckling leads to a thickness $d=1.39$ \AA, with each Co ion sandwiched between two top and two bottom Se ions. Another consequence of buckling is that the in-plane four-fold rotational symmetry is broken, which results in the inequality of the in-plane lattice constants i.e. anisotropy ($a$ = $5.577$ \AA, $b$ = $5.770$ \AA) and two kinds of Co-Se bonds with unequal lengths ($2.343$ \AA\ and $2.357$ \AA), according to our calculations. The phonon spectrum (Fig.~\ref{fig2}(b)) shows no obvious imaginary modes, confirming its dynamical stability. The elastic constants $C_\text{ij}$ have been calculated using energy-strain method by VASPKIT~\cite{VASPKIT} as shown in Table~S1~\cite{sm}. The pentagonal CoSe$_2$ monolayer meets the Born criteria: $C_{11} > 0$, $C_{66} > 0$, and $C_{11} \times C_{22} > C_{12} \times C_{12}$, indicating the mechanical stability~\cite{Long2019-ACSAEM}. Furthermore, angle-dependent Young's modulus and Poisson's ratio derived from elastic constants $C_\text{ij}$ can be found in Fig.~S3~\cite{sm}. As depicted in Fig.~S4~\cite{sm}, AIMD simulation at $300$ K has been performed to check the thermal stability of CoSe$_2$ monolayer. There is no obvious distortion after a $10$ ps AIMD simulation at $300$ K, demonstrating the thermodynamic stability of CoSe$_2$ monolayer. 

Unpaired $3d$ orbitals may exist in the pentagonal CoSe$_2$ monolayer, potentially inducing magnetism. Our DFT calculation indeed finds $1$ $\mu_{\rm B}$/Co for the ferromagnetic state, implying the low-spin state of $3d^5$ or $3d^7$. The orbital-projected density of state (PDOS) of one Co ion is shown Fig.~\ref{fig2}(c). It seems that the $t_{\rm 2g}$ orbitals ($d_{xy}$/$d_{yz}$/$d_{xz}$) are fully occupied, while the $d_{x^2-y^2}$ orbital is fully empty. The $d_{z^2}$ orbital is only occuplied by the spin-up electron. The shape of spin-polarized electron cloud also confirms the $3d^7$ scenario, i.e. Co$^{2+}$ \& Se$^{1-}$. This unusual valence can be attributed to the relatively small electronegativity difference between Co (Pauling $\chi$ = 1.88) and Se (Pauling $\chi$ = 2.55). Indeed, Se can exhibit the oxidation state of $-1$ in certain systems, such as sodium diselenide (Na$_2$Se$_2$)~\cite{KRIEF2002-TL}. The orbital splitting and filling of Co$^{2+}$ ion is summarized as the inset of Fig.~\ref{fig2}(c). 

Four magnetic configurations in a $2 \times 2$ supercell (Fig.~S5 in SM~\cite{sm}), including ferromagnetic (FM), N\'eel antiferromagnetic (N-AFM), stripe antiferromagnetic (S-AFM), and double-stripe antiferromagnetic (D-AFM) have been calculated to determine the magnetic ground state. Our calculations indicate the N-AFM state is the energetically most favorable, while the energies of FM, S-AFM, and D-AFM states are $36.9$, $21.8$, and $17.5$ meV per unit cell (u.c.) higher than the ground state. The magnetic moments originate primarily from Co$^{2+}$ ions, while the contribution of Se$^{1-}$ ions are very small. Although the electron number of each Se$^{1-}$ ion is odd, the formation of Se-dimers [Fig.~\ref{fig2}(a)] quenches their local moments via covalent bonding.

According to the energy profile obtained via calculation, the exchange interactions are extracted, including the nearest-neighbor coupling ($J_1$), second nearest-neighbor coupling ($J_2$), third nearest-neighbor coupling ($J_3$), which can be summarized in Table~S2 in SM~\cite{sm}. 

\begin{figure}
	\centering
	\includegraphics[width=0.48\textwidth]{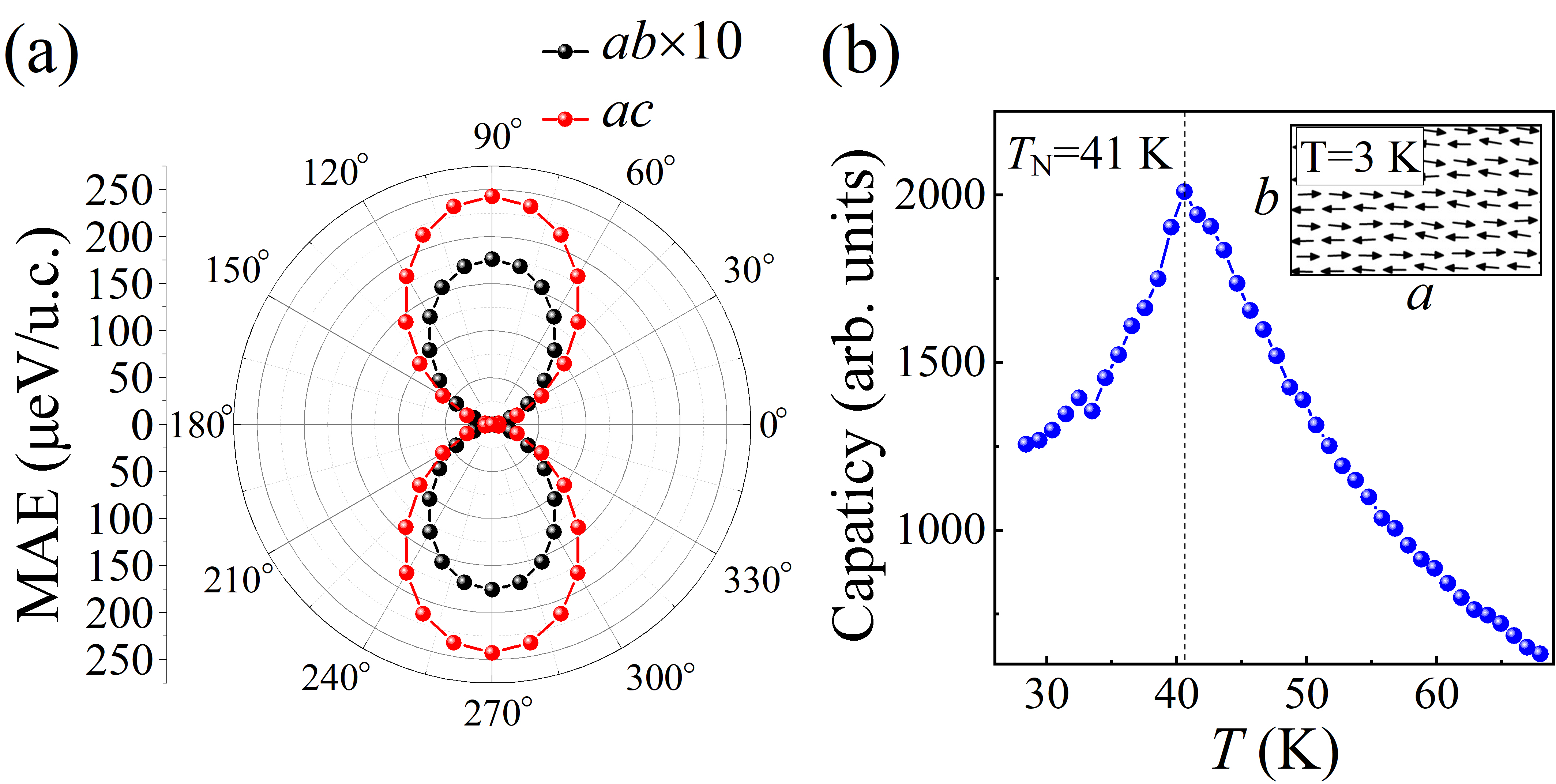}
	\caption{(a) The MAE of CoSe$_2$ monolayer as a function of spin orientation in the $ab$ and $ac$ planes. The value of $ab$-plane has been amplified tenfold for better comparison. The radial axis represents the angle difference of the spin orientation relative to the $x$-direction. And $\theta$=$0\degree$ indicates the $x$ directon in the $ab$ and $ac$ planes. (b) The MC simulation of antiferromagnetic-paramagnetic phase transition, indicated by the peak of heat capacity, which represents the total specific heat rather than per lattice site. Inset: a MC snapshot at $3$ K.}
	\label{fig3}
\end{figure}

The magnetocrystalline anisotropy of pentagonal CoS$_2$ monolayer has been calculated by including spin-orbit coupling (SOC) effect. The angle-dependent magnetic anisotropy energies (MAE) are depicted in Fig.~\ref{fig3}(a). The energy variation for in-plane ($ab$-plane) spin rotation is one order of magnitude smaller than that for out-of-plane ($ac$-plane) spin rotation. For the N-AFM state, the easy magnetization axis is along the $a$-axis, with an energy lower than that along the $b$-axis and $c$-axis for $18$ $\mu$eV/u.c. and $242$ $\mu$eV/u.c. respectively.

Then a spin model Hamiltonian can be written in the following form:
\begin{equation}
	\begin{split}
		H = & -J_{1} \sum_{\langle i, j\rangle} \mathbf{S}_{i} \cdot \mathbf{S}_{j} 
		-J_{2} \sum_{\langle\langle m, n\rangle\rangle} \mathbf{S}_{m} \cdot \mathbf{S}_{n} \\
		& -J_{3} \sum_{\langle\langle\langle k, l\rangle\rangle\rangle} \mathbf{S}_{k} \cdot \mathbf{S}_{l} 
		-A \sum_{i}\left(\mathbf{S}_{i}^{z}\right)^{2},
	\end{split}
\end{equation}
where $S$ represents the normalized spin (i.e. $|S|=1$) and $A$ represents the single-axis magnetic anisotropy term. Note that here the much weaker in-plane MAE is neglected for simplify. With this model Hamiltonian, MC simulation has been performed. The peak of heat capacity appears at $T_{\rm N}\sim41$ K [Fig.~\ref{fig3}(b)], which indicates the magnetic phase transition. The low-temperature MC snapshot confirms the N-AFM as the magnetic ground state.

\begin{figure}
	\centering
	\includegraphics[width=0.48\textwidth]{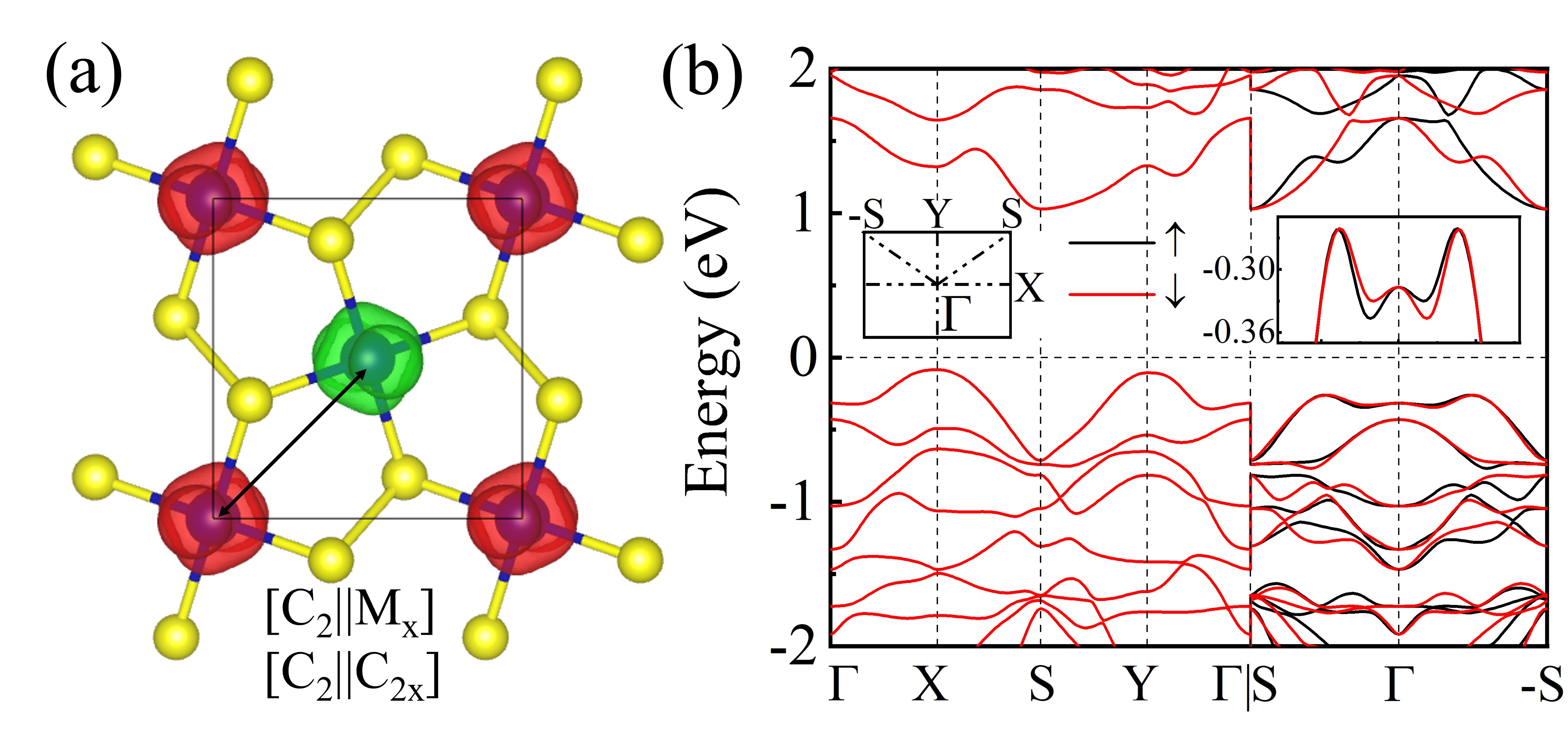}
	\caption{(a) Spin charge density and symmetry operations connecting two opposite spins of pentagonal CoSe$_2$ monolayer with N-AFM order. The red and green sites are spin-up and spin-down. (b) The electronic band structure in the first Brillouin zone. Left panel: without splitting along high symmetry $k$-paths. Inset: 2D Brillouin zone. Right panel: with splitting along other $k$-paths. Inset: magnified view around the $\Gamma$ point.}
	\label{fig4}
\end{figure}

\textit{Altermagnetism \& magnetoelastic coupling}.
To clarify the possible altermagnetism, the spin group is adopted to describe the symmetry operations. In 2D materials, four symmetry operations [$R_i||R_j$] protect spin degeneracy in non-relativistic band structure throughout the Brillouin zone, where $R_i$ and $R_j$ represent the spin and spatial symmetry operations, respectively. These operations are [$C_2||\tau$], [$C_2||\bar{E}$], [$C_2||m_z$], and [$C_2||C_{2z}$]. CoSe$_2$ monolayer possesses a nontrivial spin layer group $^22/^2m_x$ that lacks these four symmetry operations, and thus the spin splitting is expectable. The relevant symmetry operations for this system can be described as:
\begin{equation}
	[E||\mathcal{P}]+[C_2||M_x,C_{2x}]
\end{equation}
where $E$ represents the identity operation for spins, which links the sites with same spin. $C_2$ denotes the antisymmetric spin flip operation that links the sites with opposite spin (Fig.~\ref{fig4}(a)). Operators $\mathcal{P}$, $M_x$, and $C_{2x}$ indicate the space inversion, mirror symmetry, and twofold rotation symmetry, respectively. For high-symmetry lines $\Gamma$-Y ($k_x=0$) and $\Gamma$-X ($k_y=0$ ), the symmetry operation $[C_2||M_x]$ and $[C_2||C_{2x}]$ ensures the Kramers spin degeneracy, respectively. Thereby, one have spin degenerate bands: 
\begin{equation}
	\begin{split}
	[C_2||M_x] E(s, 0, k_y)=E(-s,0, k_y), \\
	[C_2||C_{2x}]E(s,k_x, 0)=E(-s,k_x,0).
	\end{split}
\end{equation}
Consequently, one can obtain:
\begin{equation}
	\begin{split}
	E(s, 0, k_y)=E(-s,0, k_y), \\
	E(s,k_x, 0)=E(-s,k_x,0).
	\end{split}
\end{equation}
This explains the spin degeneration along the high-symmetry lines $\Gamma$-X and $\Gamma$-Y. While for other general $k$-paths, e.g. $\Gamma$-S, the band structure exhibits alternating momentum-dependent spin splitting,  as depicted in Fig.~\ref{fig4}(b). Notably, the spin-splitting energy at the top of valence band can reach up to $213$ meV, comparable to other 2D altermagnetic materials like VBr$_4$ ($150$ meV) and RuF$_4$ ($\sim 240$ meV) \cite{Jin2024-PRB,Sodequist2024-APL,Wang2022-PRB}. 

\begin{figure}
	\centering
	\includegraphics[width=0.48\textwidth]{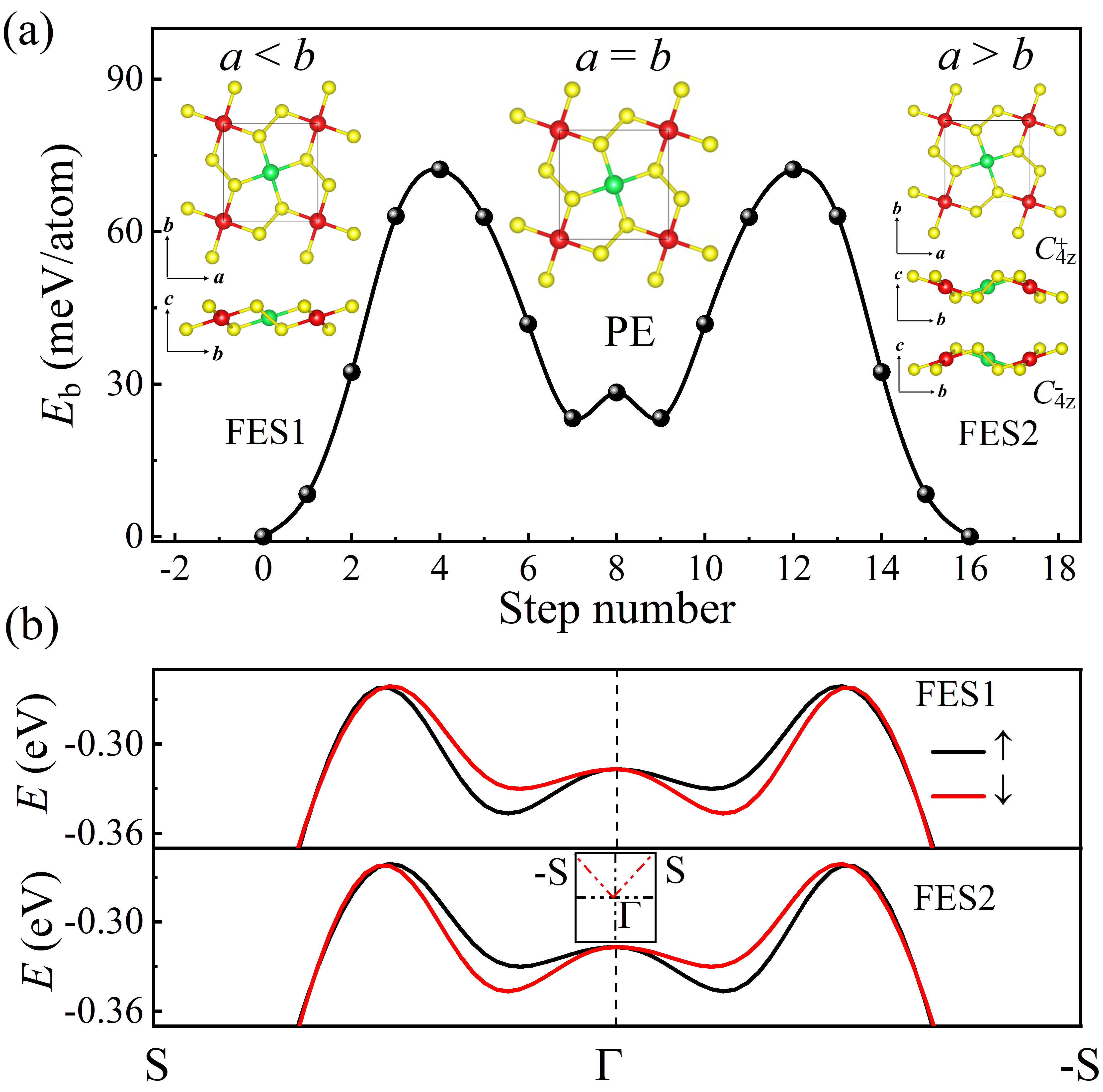}
	\caption{(a) The energy barrier of ferroelastic switching estimated by the CI-NEB method. The insets illustrate the two states before ($a$<$b$) and after ($a$>$b$) ferroelastic switching, along with a hypothetic intermediate state ($a$=$b$). The red (green) atoms represent spin-up (spin-down) magnetic moments. (b) Spin-resolved band structures of the top of valence band for the FES1 and FES2 states without spin-orbit coupling. Inset: the first Brillouin zone.}
	\label{fig5}
\end{figure}

Besides the aforementioned intriguing magnetic properties, pentagonal CoSe$_2$ monolayer also exhibits ferroelastic and magnetoelastic behavior due to the buckling effect. Ferroelastic materials own two or multiple energetically degenerate states that undergo reversible switching between variants under external mechanical strain. Here the two degenerate states can be defined as the structures with lattice constants $a<b$ (FES1) and $a>b$ (FES2), as shown in Fig.~\ref{fig5}(a). These two ferroelastic states can be connected via one of two symmetry operations, i.e. the $90\degree$ clockwise or counterclockwise rotations ($C_{4z}^{+}$ {\it vs} $C_{4z}^{-}$), which can be described mathematically as:
\begin{equation}
C_{4z}^{+} = \begin{pmatrix} 0 & -1 & 0 \\ 1 & 0 & 0\\ 0 & 0 & 1 \end{pmatrix},  \quad C_{4z}^{-} = \begin{pmatrix} 0 & 1 & 0 \\ -1 & 0 & 0\\ 0 & 0 & 1 \end{pmatrix}.
\end{equation}

The $C_{4z}^{+}$ and $C_{4z}^{-}$ operations lead to two different crystalline structures, which can be interconnected via $\mathcal{M_\text{z}}$ symmetry operation. The key distinction lies in the positions of the Se atoms, whereas the magnetic order on the two Co ions remains identical (right insets of Fig.~\ref{fig5}(a)). Thus, the spin-polarized band structures of two states of FES2 ($C_{4z}^{+}$ and $C_{4z}^{-}$) are also identical. Being energetically degenerate, they show identical switching barriers, as illustrated in Fig.~\ref{fig5}(a). The hypothetic paraelastic state (PES) occurs at $a=b=5.673$ \AA. The climbing image-budged elastic band (CI-NEB) method has been performed to simulate the ferroelastic switching process. The calculated energy barrier ($E_\text{b}$) is $\sim72$ meV/atom, which is comparable to that of AgF$_2$ ($51$ meV/atom) \cite{Xu2020-NH}, but much lower than those of Nb$_2$SiTe$_4$ ($\sim237$ meV/atom) \cite{Zhang2020-JPCL} and phosphorene ($\sim200$ meV/atom) \cite{Wu2016-NL}.

The ferroelastic strain in CoSe$_2$ monolayer can be characterized by the anisotropy, determined as $a/b-1$, which is $3.3\%$ for the FES1 state. This value is comparable to that of typical 2D ferroelastic material SnSe ($2.1\%$) \cite{Wu2016-NL}.

The ferroelastic switching involves a $90\degree$ rotation of the crystal, which correspondingly rotates both the band structure and associated spin splitting features, as depicted by Fig.~S6 in SM \cite{sm}. As specifically illustrated in Fig.~\ref{fig5}(b), along non-high-symmetry paths, the ferroelastic switching can also switch the altermagnetic spin-splitting, which would induce changes in other physical properties.

In altermagnets, spin-polarized current can be generated by electric field due to spin-splitting, which can be characterized as spin-resolved conductivity tensor $\sigma_{ij}^{s}$ ($s$ is $\uparrow$, $\downarrow$ for spin index)~\cite{Gonz2021-PRL,Dou2025-PRB,Ma2021-NC}. As depicted in Fig.S7~\cite{sm}, the diagonal components of the conductivity are spin-degenerate ($\sigma_{ii}^{\uparrow}$=$\sigma_{ii}^{\downarrow}$, $i=x,y$), whereas the off-diagonal component is spin-polarized ($\sigma_\text{xy}^{\uparrow}\neq\sigma_\text{xy}^{\downarrow}$). In addition, ferroelasticity can reverse the sign of off-diagonal spin-polarized conductivity $\sigma_\text{xy}$. Namely, $\sigma_\text{xy}^{\uparrow}$ and $\sigma_\text{xy}^{\downarrow}$ change in opposite directions. The anisotropic spin polarization can be defined as $SP=\frac{\sigma_{ij}^{\uparrow}-\sigma_{ij}^{\downarrow}}{\sigma_{ij}^{\uparrow}+\sigma_{ij}^{\downarrow}}=\frac{2\sigma_\text{xy}^{\uparrow}sin(2\varphi)}{\sigma_\text{xx}cos^2\varphi+\sigma_\text{yy}sin^2\varphi}$, where $\varphi$ denotes the angle between the direction of the electric field and the $x$-axis. And we further observe that ferroelasticity also reverse its sign. These results indicate the ferroelasticity can effectively manipulate spin-polarized currents, enabling the design of strain-controlled spintronic devices for information encoding and sensing.

\begin{figure}
	\centering
	\includegraphics[width=0.48\textwidth]{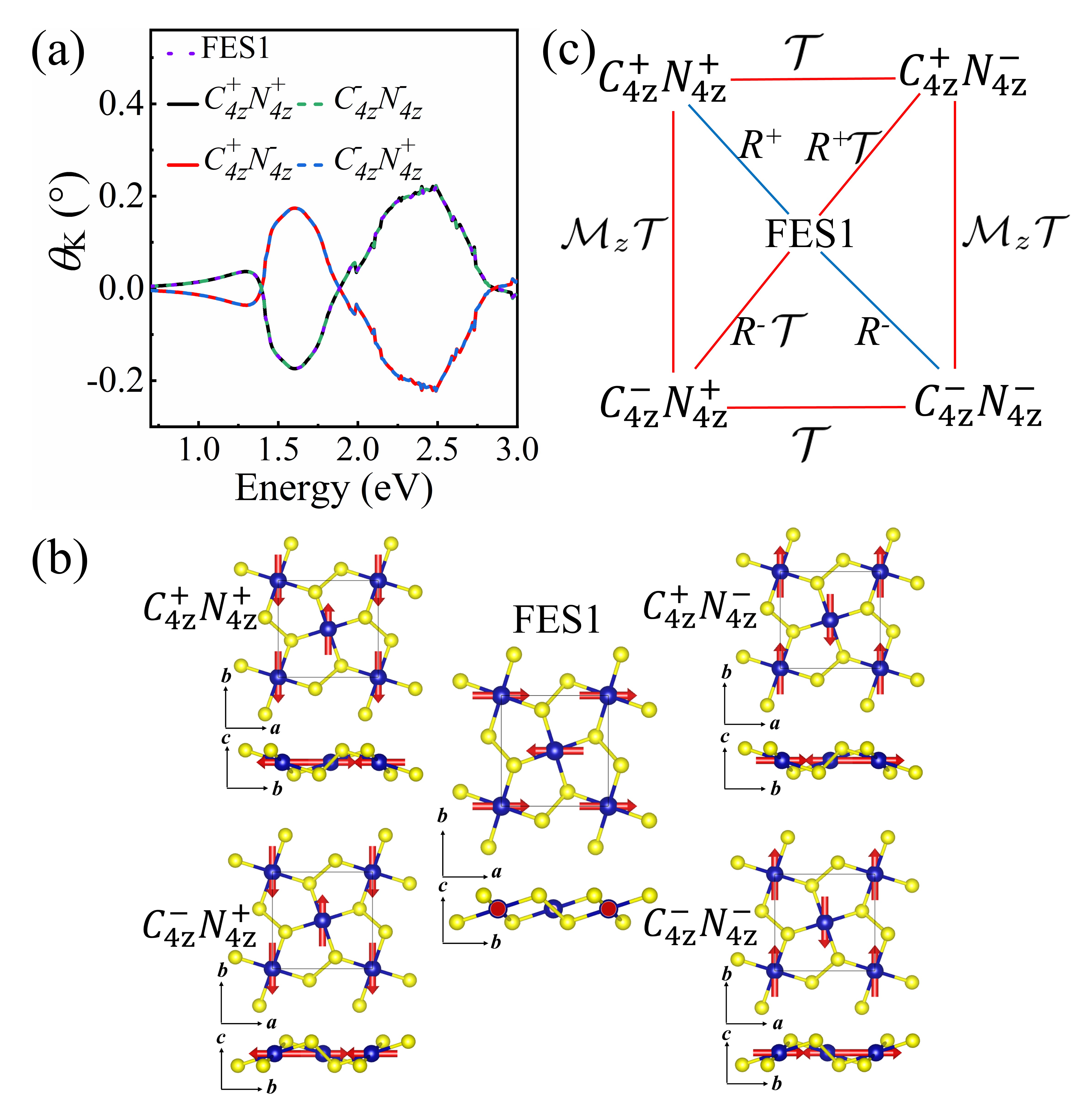}
	\caption{(a) Energy-dependent magneto-optical Kerr spectra  of different states inferred from Wannier90 calculations. (b) FES1 state and four possible states emerging after ferroelastic switching when taking SOC into consideration. $C_{4z}$ and $N_{4z}$ indicate crystal and spin rotate $90\degree$ respectively, then the positive (+) and negative (-) signs control clockwise and counterclockwise rotation respectively. (c) Schematic diagram of symmetry operations connecting different states, where red (bule) lines indicate pairs of states with opposite (identical) magneto-optical Kerr angles.} 
	\label{fig6}
\end{figure}

\textit{Switching of magnetic-optical Kerr effect}.
Taking spin-orbit coupling into consideration, MOKE may be occured in altermagnets. For CoSe$_2$ monolayer with the N-AFM order, the corresponding magnetic space group (MSG), magnetic point groups (MPG), and optical conductivity tensor $\xi_{ij}$ can be found in Table~S3 in SM \cite{sm}. The MOKE response in this system exhibits strong N\'eel vector orientation dependence. When the magnetization direction aligns along the $x$-axis, there are two nonzero off-diagonal components of optical conductivity tensor, namely $\xi_\text{xy}$ and $\xi_\text{yz}$. However, due to the 2D nature, only $\xi_\text{xy}$ survives. When the magnetization direction turns to $y$-axis or $z$-axis, MOKE becomes forbidden. Since the easy magnetization direction of CoSe$_2$ monolayer is along the $x$-axis as established earlier, the MOKE can be activated spontaneously. As shown in Fig.~\ref{fig6}(a), the Kerr angle can reach up to $0.2^{\degree}$, which is readily detectable experimentally.

Considering the direction and phase of N\'eel vector $\textbf{\textit{N}}$, defined as the difference of the magnetic moments $\textbf{\textit{M}}$ between Co$_1$ and Co$_2$ in the unit cell, i.e. $\textbf{\textit{N}}$=$\textbf{\textit{M$_{Co_1}$}}$-$\textbf{\textit{M$_{Co_2}$}}$. And $N_{4z}^{+}$ ($N_{4z}^{-}$) can describe the $90\degree$ clockwise (counterclockwise) rotation of the N\'eel vector. Thus four possible energy-degenerate FES2 states that can be derived from the FES1 state, labeled as $C_{4z}^{+}N_{4z}^{+}$, $C_{4z}^{+}N_{4z}^{-}$, $C_{4z}^{-}N_{4z}^{+}$, and $C_{4z}^{-}N_{4z}^{-}$, as shown in Fig.~\ref{fig6}(b). Here $C_{4z}$ and $N_{4z}$ represent the $90\degree$ rotations of crystal axes and N\'eel vectors respectively, and the superscripts $+$ and $-$ indicate the clockwise and counterclockwise rotations. Interestingly, these four states exhibit equal occurrence probability during the ferroelastic switching, but they contribute distinct MOKE responses, as depicted in Fig.~\ref{fig6}(a). Our analysis reveals that the MOKE response remains unchanged compared to the FES1 state when N\'eel and crystal vectors rotate homodirectionally, but the sign of MOKE signal is reversed when their rotations oppose each other. 

In most cases, the inversion of MOKE signal can be achieved via the $180\degree$ spin rotation. Here this inversion may be done via the $90\degree$ spin rotation, with the help of ferroelasticity. This phenomenon can be fully explained by symmetry analysis as depicted in Fig.~\ref{fig6}(c). $R^+$ ($R^-$) represents a $90\degree$ cooperative clockwise (counterclockwise) rotation of both crystal axes and N\'eel vectors, preserving the Kerr angle sign. In contrast, time reversal operation ($\mathcal{T}$) reverses the Kerr angle according to Onsager’s relation~\cite{Rathgen2005-PRB}. Therefore, the homodirectional rotation of the lattice and N\'eel vectors maintains the Kerr angle sign, while opposite-directional rotation reverses it, establishing a distinct magnetoelastic coupling mechanism.

In summary, based on first-principles calculations, the electronic and magnetic properties of pentagonal CoSe$_2$ monolayer have been revealed, which exhibits a N\'eel-type antiferromagnetic ground state with $T_{\rm N}=41$ K. Notably, prominent spin splitting ($214$ meV) in the absence of SOC confirms its altermagnetic nature. Remarkably, the observed ferroelastic phase transition may induce a concomitant $90\degree$ rotation of the spin splitting bands. Accompanying the ferroelastic switching, the homodirectional rotation of the lattice axes and N\'eel vectors can preserve the Kerr angle sign, while the opposite-directional rotation inverts it, unveiling a novel magnetoelastic coupling mechanism. Our work not only establishes a fundamental approach for manipulating altermagnetic states in more general multiferroic systems but also paves the way for investigating unconventional magenetoelastic phenomena. Importantly, these findings are generalizable to analogous systems, including RuF$_4$, V$X_4$ ($X$ = F, Br, I) and MnF$_4$ monolayers.

\textit{Note added.} Recently, we have noticed two works on ferroelastic altermagnets~\cite{Guo2025-PRB,Huang2025-PRL}. The work by Guo \textit{et al}. coincides with our findings, demonstrating that ferroelasticity can reverse the spin-splitting in altermagnets~\cite{Guo2025-PRB}. The study by Huang \textit{et al}. also emphasizes the modulating effect of ferroelasticity on the band structure of altermagnetic systems, which can induce a transition from an altermagnetic to a compensated ferrimagnetic state~\cite{Huang2025-PRL}.

\begin{acknowledgments} We thank Professor Shengyuan A. Yang for his insightful and helpful discussions. This work was supported by the National Natural Science Foundation of China (Grants No. 12274069, No. 12325401 and No. 12104089), Jiangsu Funding Program for Excellent Postdoctoral Talent under Grant Number 2024ZB001, China Postdoctoral Science Foundation under Grant Number 2024M760423, the Fundamental Research Funds 356 for the Central Universities (Grant No. 2242025K30023) and the Big Data Computing Center of Southeast University.
\end{acknowledgments}
\bibliography{cite}

\end{document}